\documentclass[12pt]{article}

\usepackage{a4}
\usepackage[latin1]{inputenc}
\usepackage{openbib}
\usepackage{epsfig, rotating}
\usepackage{graphicx}
\usepackage{subfigure}
\usepackage{latexsym}
\usepackage{amssymb}
\begin{document}

\setlength{\parskip}{1.5ex}

\thispagestyle{empty}


\vspace*{1cm}

\begin{center}
{\Large \bf An Improved Weighting Algorithm to Achieve Software
Compensation in a Fine Grained LAr Calorimeter}

\vspace*{2cm}

{\large \bf  \c{C}. \.{I}\c{s}sever\footnote{Present address:
University of California, Santa Barbara}, K. Borras\footnote{Present
address: DESY, Hamburg}, D. Wegener\\ Institute of Physics,
University of
Dortmund}\\

\end{center}

\vspace*{2cm}

\underline{Abstract}\\ An improved weighting algorithm applied to
hadron showers has been developed for a fine grained LAr
calorimeter. The new method uses tabulated weights which depend on
the density of energy deposited in individual cells and in a
surrounding cone whose symmetry axis connects the interaction
vertex with the highest energy cluster in the shower induced by a
hadron. The weighting of the visible energy and the correction for
losses due to noise cuts are applied in separate steps. In
contrast to standard weighting procedures the new algorithm allows
to reconstruct the total energy as well as the spatial energy
deposition on the level of individual calorimeter cells.

The linearity and the energy resolution of the pion signal in the
momentum interval $2~ GeV/c \leq p \leq 20 ~GeV/c$ studied in this
 analysis are considerably improved in comparison to the standard
weighting algorithm as practiced presently by the H1
collaboration. Moreover the energy spectra reconstructed with the
new method follow in a broad interval a Gaussian distribution and
have less pronounced tails.

\newpage

\section{Introduction}

The different response of hadron calorimeters to electromagnetic
and hadronic showers, observed for the first generation of hadron
calorimeters \cite{Engler1, Engler2, Moritz}, are a source of
non-linearities, deviations of the relative resolution from the
statistically expected $1/\sqrt{E}$ -behaviour and the origin of
the non-gaussian shape of the response function \cite{CDHS,
Wigmans}. Two different methods have been proposed to avoid these
disadvantages. Hardware compensation, optimized by extensive
simulations of hadron showers
 \cite{Wigmans, Gabriel, Brueckmann},
enhances the hadron signal by detecting neutrons, produced in
hadron $^{238}U$ interactions, with scintillators \cite{ZEUS,
WA78}. On the other hand in fine grained calorimeters the signal
of different detector cells can be weighted in such a way that the
overall signal of electrons and hadrons depositing the same energy
is equalized. This method was pioneered by the CDHS Collaboration
\cite{CDHS} using a large scintillator-iron calorimeter to detect
hadron showers produced in $\nu$-events. Essential improvements
were achieved by the H1 Collaboration \cite{H11, H11c, H12, H13}
exploiting the high granularity of its 45000 channel LAr
calorimeter.

In this paper a new weighting algorithm is presented which not
only allows to determine the deposited energy in an improved way
but in addition permits a more realistic reconstruction of the
spatial shower distribution. While previous algorithms aimed just
for a determination of the total shower energy, the new method
intends to reconstruct the energy in each cell of the calorimeter
hit by the hadronic shower. Losses due to noise suppression cuts
applied to individual cells are considered separately by adding a
correction term to the total reconstructed energy.

After a short description how the visible energy is derived from
the measured charge, which is identical to the standard procedure
of the H1 algorithm \cite{Goerlich, Wellisch, Shekelyan}, the new
method is explained. The weights, and the measurable variables
they depend on, are determined next, followed by a description of
the algorithm, correcting for losses due to noise. Finally the new
weighting algorithm is applied to Monte Carlo and test beam data.
The results achieved are compared to those of the standard H1
algorithm described in \cite{Goerlich, Wellisch, Shekelyan}.

\section{Reconstruction of the Visible Energy}

\subsection{Determination of the Electromagnetic Calibration Constant}

The electromagnetic calibration constant which allows to convert
the recorded charge into visible energy, has been derived by an
iterative procedure \cite{Gay} equalizing the reconstructed energy
of test beam and simulated data for electrons:
\begin{equation}
  \large \langle E^{exp}_{0}\rangle =  \large \langle E^{sim}_{0}\rangle
\label{eq1}\end{equation}

The experimentally determined energy is derived  from the
deposited charges $Q_{i}$

\begin{equation}
  \large \langle E^{exp}_{0}\rangle = C_{exp} \langle ~\sum_{i}~Q_{i}\rangle
\label{eq2}\end{equation}

where the sum runs over all calorimeter cells with a signal
passing the noise cuts. The reconstructed signal of simulated data
is given by the expression

\begin{equation}
  \large \langle E^{sim}_{0}\rangle =  \langle
 ~ \sum_{i}~(C_{sim}~E_{vis}^{i} + C_{exp}~Q_{noise}^{i}) ~\rangle
\label{eq3}\end{equation}

\begin{equation}
 \large C_{sim} =  \langle \frac{E_{dep}}{E_{vis}} \rangle
\label{eq4}\end{equation}

$C_{sim}$ is derived from simulated data. $E_{dep}$ is the energy
deposited in the calorimeter, while

\begin{equation}
E_{vis} = \sum_{i}~E^{i}_{vis} \label{eq4a}\end{equation}

is the detected visible energy in the active medium of the
calorimeter derived by simulation. The influence of saturation
phenomena, charge losses due to electron capture by
electronegative gases \cite{Hof1}, recombination in the ionization
column \cite{Jaffe} etc. was derived from HV curves recorded
before and after the data collection. In addition probes similar
to those used finally in the H1 calorimeter \cite{Barre} monitored
the response to the signal of a radioactive source continuously.
$Q^{i}_{noise}$ is the measured noise in calorimeter channel $i$.
$C_{exp}$ is obtained iteratively from eq. (\ref{eq2}) -
(\ref{eq4}) requiring the constraint (\ref{eq1}) to hold for
electrons.

The calibration constant $C_{exp}$ determined by this method
defines the electromagnetic scale of the calorimeter. The
reconstructed energy $E_{0}^{i}$ of cell $i$ on the
electromagnetic scale for test beam data is given by the
expression

\begin{equation}
E^{i}_{0} = C_{exp}~Q_{i} \label{eq5}\end{equation}

and for simulated data by

\begin{equation}
 E^{i}_{0} = C_{sim} ~E^{i}_{vis} + C_{exp}~Q^{i}_{noise}
\label{eq6}\end{equation}

in analogy to eqs. (\ref{eq2}) and (\ref{eq3}).

The data used in the present analysis were collected at the CERN
SPS test beam H6 \cite{Korn} using a FB2-type module of the H1 LAr
calorimeter \cite{H1 Det, H196}. The simulation employed the H1
software packages H1SIM \cite{H1SIM} and ARCET \cite{ARC} based on
the event generators GEANT~3.21 \cite{Brun} and
GHEISHA\cite{Fesefeldt}.

\subsection{Clustering of Energy}

The identification of calorimeter cells hit by the hadron showers
follows the standard methods developed by the H1 Collaboration,
described in more detail in ref. \cite{Goerlich, H194, Shekelyan}.
Besides the charge produced by the hadron shower electronic noise
contributes to the detected signal which amounts typically to
$\sigma_{noise} = 15~MeV$ up to $30~MeV$ per channel \cite {H196}.
Only charges with

\begin{equation}
|Q_{i}|~>~2.5 \cdot \sigma_{noise}^{i} \label{eq7}\end{equation}

are recorded for the calorimeter modules used in the present
analysis. In the analysis however an additional cut is applied.
Besides those channels with

\begin{equation}
|Q_{i}|~\geq~4~ \sigma_{noise}^{i} \label{eq8}\end{equation}

in addition cells are taken into account which are direct
neighbours of a channel with $Q_{i}~>~4~\sigma^{i}_{noise}$:

\begin{equation}
Q_{i} > 4~ \sigma_{i}~ \wedge~|Q_{j}|~ > 2.5~ \sigma_{j}~
\mbox{with}~ j = \mbox{neighbour cell of}~ i
\label{eq8.1}\end{equation}

Clusters combining those cells which pass the noise cuts
(\ref{eq7}), (\ref{eq8}), (\ref{eq8.1}) are constructed in two
steps \cite{Goerlich}. First all cells in a given plane of the
calorimeter at constant distance from the beam are grouped into
two-dimensional clusters around the cell with the highest charge
deposit. In the second step adjacent two-dimensional clusters are
combined to a three-dimensional cluster. The topological nearest
cells to a 3D-cluster with a negative signal are added to the
original cluster. If the total charge of this combination is
negative the 3D-cluster is excluded from the following analysis.

Clusters due to primary photons and electrons are identified by
estimators which exploit the characteristic shape of an
electromagnetic shower \cite{H194, Wellisch}. They are not
considered further in the analysis. The remaining clusters are
labeled as hadronic clusters, all weights are derived from and
applied to this subset of the data.

The quantity $E_{cone}$, the energy deposited inside a cone whose
axis connects the interaction vertex with the most significant
hadronic cluster, where its significance is defined by

\begin{equation}
p = \sqrt{ \sum_{i} \left( \frac{E_{0}^{i}}{\sigma^{i}_{noise}}
\right)^{2}} \label{eq9}
\end{equation}

turns out to be an important input variable for the weighting
algorithm developed in this paper. $E_{cone}$ is determined in two
steps following essentially the procedure developed for the H1
standard analysis \cite{Wellisch}. If no other cluster apart from
the one defining the cone axis is found within an opening angle of
$11^{o}$ the energy of the cone is given by

\begin{equation}
E_{cone} = E_{cluster} \label{eq10}
\end{equation}

However, if more than one cluster is recorded within this cone, a
new one of $11^{o}$ opening angle is constructed whose axis
connects the interaction vertex with the energy center of gravity
of the hadron clusters of the previous cone. In this case the cone
energy is defined by the sum over all clusters in the new cone

\begin{equation}
E_{cone} = \sum_{j}~E^{j}_{cluster} \label{eq11}\end{equation}

In the following the cone reconstructed with the largest energy is
referred to as "most energetic cone"; $E^{max}_{cone}$ denotes its
energy. Further cones are constructed from the remaining hadron
clusters following the procedure described above.

\subsection{Description of the New Weighting Algorithm}

Two components of the hadron showers strongly influence the energy
deposition and the possibilities to measure its energy. The break
up of nuclei strongly reduces the detectable energy \cite{Wigmans,
Gabriel, Brueckmann}; the lost energy is referred to as "invisible
energy" in the following, it has to be compensated by weighting.
On the other hand the electromagnetic component in a hadron shower
is deposited in the calorimeter without losses and therefore has
not to be weighted.

The energy density

\begin{equation}
\rho_{i} = \frac{E^{i}_{0}}{Vol^{i}}
 \label{eq19}\end{equation}

in cell $i$ with volume $Vol^{i}$ allows to tag these components
 \cite{CDHS}.

\begin{figure}[h!]
  \begin{center}
   \includegraphics[angle=0, width=7.5cm]{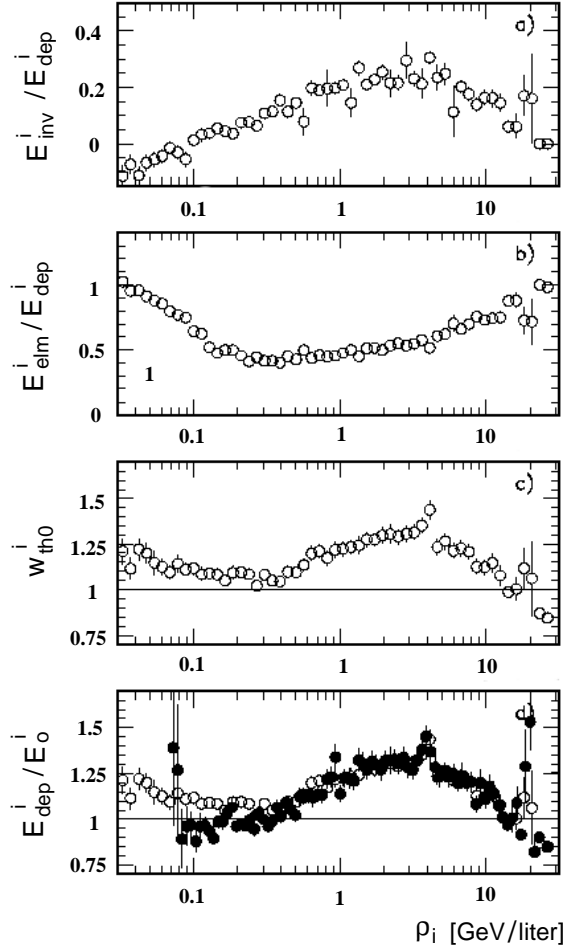}
    \end{center}
 \caption{Fraction of energy deposited in cell $i$ normalized to the corresponding
reconstructed energy as function of the energy density $\rho_i$:
a) invisible energy, b) electromagnetic energy. Theoretical
weights $w^{i}_{th, o}$ according to eq. (\ref{eq12a}) are plotted
as a function of $\rho_i$ in c) and are compared to the weights
$w^{i}_{th}$ (full points) of eq. (\ref{eq12}) in d). The energy
of the primary hadron is 15 GeV. }
 \label{abb1}
\end{figure}

In figs. \ref{abb1}a, b the fractional contribution of the
invisible energy and the electromagnetic energy to the total
deposited energy in cell $i$ respectively are plotted as functions
of the energy density $\rho_{i}$. The electromagnetic component
rises with increasing energy density. This behaviour can be traced
back to the fact that electromagnetic subshowers have a smaller
spatial extension than hadron induced subshowers.

In fig.\ref{abb1}c the theoretical weights

\begin{equation}
w^{i}_{th,o} = \frac{E^{i}_{dep}}{C_{sim}~E_{vis}^{i}}
\label{eq12a}\end{equation}

which according to eq.(\ref{eq6}) do not consider the noise
contribution are shown as a function of $\rho_{i}$. It reveals a
strong variation with $\rho_{i}$ resulting from the individual
dependencies of the invisible and the electromagnetic component on
the energy density. $w^{i}_{th,o}$ is compared in fig.\ref{abb1}d
to the weight

\begin{equation}
w^{i}_{th} =  \frac{E^{i}_{dep}}{ E^{i}_{0} }
\label{eq12}\end{equation}

which allows to convert the reconstructed energy $E^{i}_{o}$ of
eq.(\ref{eq6}) on the electromagnetic scale into the real
deposited energy $E^{i}_{dep}$. The differences $w^{i}_{th}$ and
$w^{i}_{th,o}$ at small energy densites $\rho_{i}$ can be
explained qualitatively as a consequence of the overlayed noise
and the noise cuts applied \cite{CI}.

The theoretical weights $w^{i}_{th,o}$ in the interval $3~
\frac{GeV}{l} \leq \rho_{i} \leq 30~\frac{GeV}{l}$ decrease
(fig.\ref{abb1}c), since in this region the relative contribution
of nuclear binding energy diminishes (fig.\ref{abb1}a), i.e. less
energy has to be corrected for. In the interval $0.2~
\frac{GeV}{l} \leq \rho_{i} \leq 3~\frac{GeV}{l}$ the theoretical
weights $w^{i}_{th,o}$ increase with $\rho_{i}$, this can be
attributed to the growth of the invisible energy
(fig.\ref{abb1}a). Especially at lower values of $\rho_{i}$ the
energy of charged particles is deposited by excitation and
ionization of atoms; hence contributions from nuclear reactions
can be neglected. In the interval $\rho_{i} < 0.2~\frac{GeV}{l}$
the weight $w^{i}_{th,o}$ increases with decreasing $\rho_{i}$.
Since these energy densities are characteristic for energy
depositions of low energy electrons and photons in the tail of the
shower (see increase of the fractional contribution of the
electromagnetic energy in fig.\ref{abb1}b), the transition effect
\cite{Wigmans, Brueckmann, ARGUS} reduces the signal and therefore
forces the weights to increase.

\begin{figure}[h!]
  \begin{center}
   \includegraphics[angle=0, width=8cm]{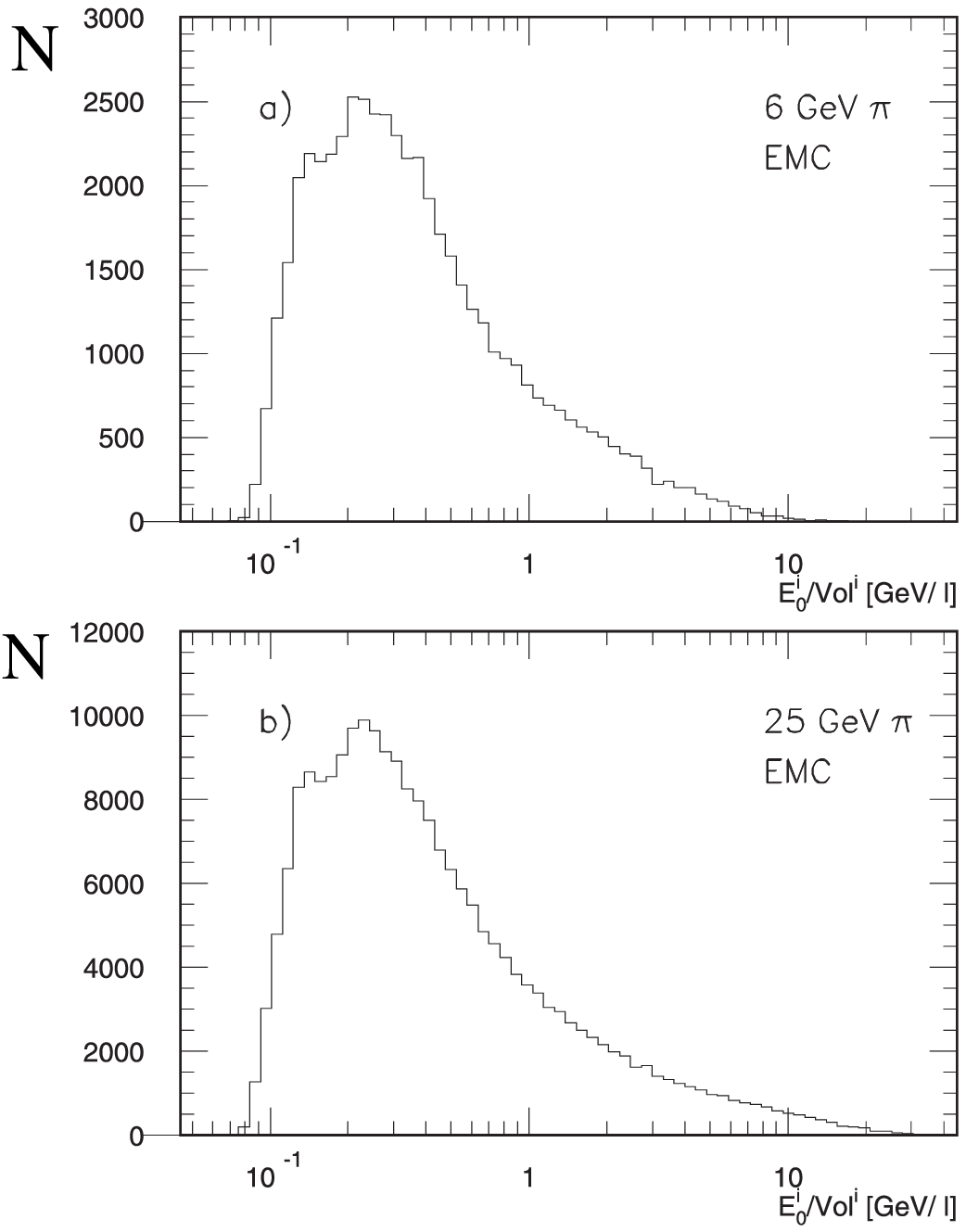}
    \end{center}
 \caption{Distribution of energy density of cells in the "most energetic cone" for
pions of a) 6 GeV/c and b) 25 GeV/c.}
 \label{abb2}
\end{figure}

Fig.\ref{abb2} demonstrates that in first approximation the shape
of the energy density distribution is independent of the energy of
the primary pion, but a closer look reveals differences especially
at large values of $\rho_{i}$. This can be attributed to the
increase of the electromagnetic fraction in hadron showers at
higher primary energies \cite{Gabriel2}. This effect has to be
taken into account in a global way; the energy $E_{cone}$ of the
cone provides the necessary information to estimate the initial
hadron energy \cite{CI} as will be demonstrated.

Hence the weighting factors of the new algorithm are parameterized
as function of $\rho_{i}$ and $E_{cone}$

\begin{equation}
w^{i}_{new}~ (\rho_{i}, E_{cone}) = \left. \frac{\left<
E^{i}_{dep} \right>}{\left< E^{i}_{o} \right>} \right|_{\rho_{i},
E_{cone}} \label{eq13}\end{equation}

\begin{figure}[h!]
  \begin{center}
   \includegraphics[angle=0, width=9cm]{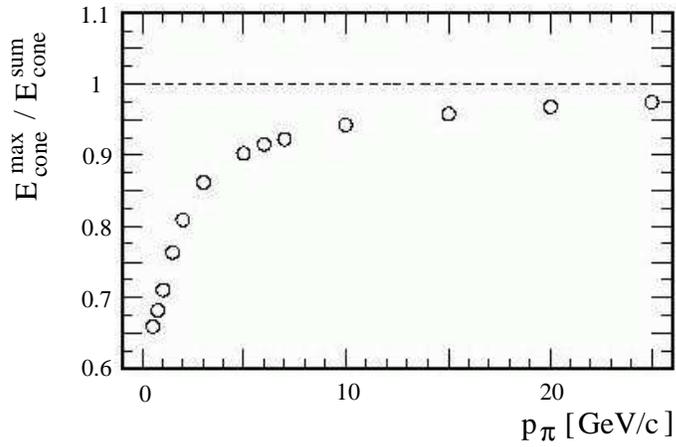}
    \end{center}
 \caption{Fraction of energy reconstructed in the "most energetic cone" normalized to
the total reconstructed energy.}
 \label{abb3}
\end{figure}

\begin{figure}[h!]
  \begin{center}
   \includegraphics[angle=0, width=8cm]{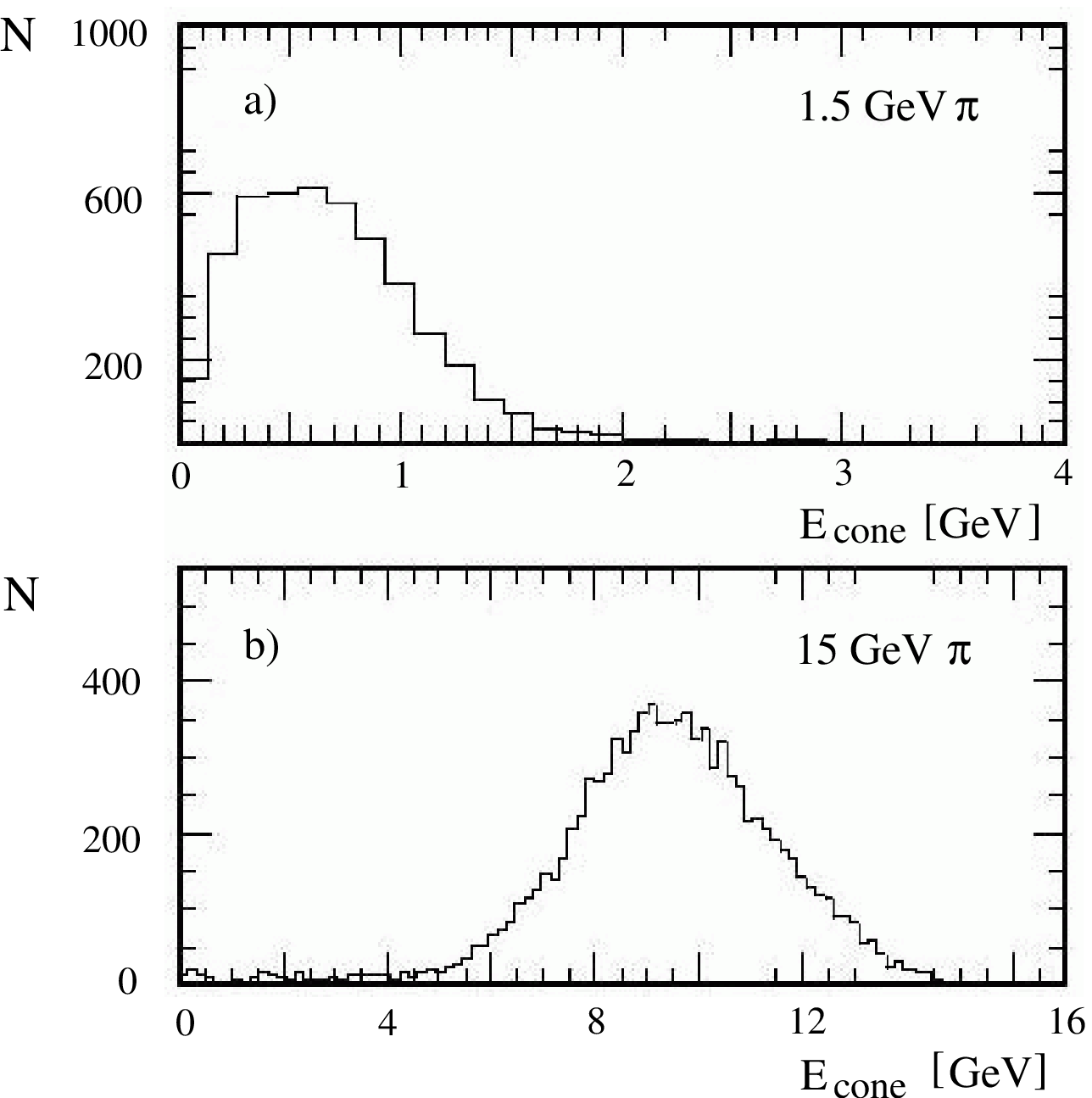}
    \end{center}
 \caption{Distribution of reconstructed energy in the "most energetic cone" a) for low
and b) for high momentum pions.}
 \label{abb4}
\end{figure}

They are derived from simulated data taking noise into account and
applying noise cuts. $\left<E^{i}_{dep}\right>$ and
$\left<E^{i}_{0}\right>$ are the mean values of the deposited and
reconstructed energy in a $\rho_{i} - E_{cone}$ interval
respectively. Only cells of the "most energetic cone" of an event
are considered, since $E^{max}_{cone}$ is a convenient measure of
the hadron energy initiating the shower. This becomes evident from
fig.\ref{abb3} where the fraction of energy deposited in the "most
energetic cone" is plotted as a function of the primary hadron
energy. For $p_{\pi}
> 2~ GeV/c$ more than $75 \%$ of the total reconstructable energy
is detected in this cone, this fraction rises strongly with
increasing $p_{\pi}$ and saturates at $\sim~95\%$. Hence for the
construction of the weights the choice of $E^{max}_{cone}$ as the
second variable, on which $w^{i}_{new}$ depends, makes sense. If
the weights of cells for cases with $E_{cone} < E^{max}_{cone}$
would have been considered in addition, corrections for showers
with low and high deposited energy would be mixed up, while
considering only the "most energetic cone" provides clean
conditions with small overlap of low and high energy data
(fig.\ref{abb4}).

The weights according to equation (\ref{eq13}) of the new
algorithm for the electromagnetic and hadronic modules of the
H1-LAr calorimeter in the test set-up are collected in figs.5a, b.
A double logarithmic scale has been chosen to depict the strong
variation of $w^{i}_{new}$ at small $\rho_{i},~E_{cone}$
\cite{CI}. The new weighting algorithm allocates for each cell $i$
according to its energy density and the energy of its assigned
cone the weighting factor $w^{i}_{new}~(\rho_{i},~E_{cone})$ from
the tables shown in figs.5. It reconstructs the energy on the
hadronic scale $E^{i}_{rec}$ by multiplying the energy on the
electromagnetic  scale $E^{i}_{o}$ with the new weight

\begin{equation}
E^{i}_{rec} = E^{i}_{0} \cdot w^{i}_{new}~(\rho_{i},~E_{cone})
\label{eq13a}\end{equation}

Note that the reconstruction of the energy on the hadronic scale
of cells, which do not belong to the most energetic cone, is also
performed with the weighting factor shown in fig.5.

\begin{figure}[h!]
  \begin{center}
  \includegraphics[angle=0, width=7cm]{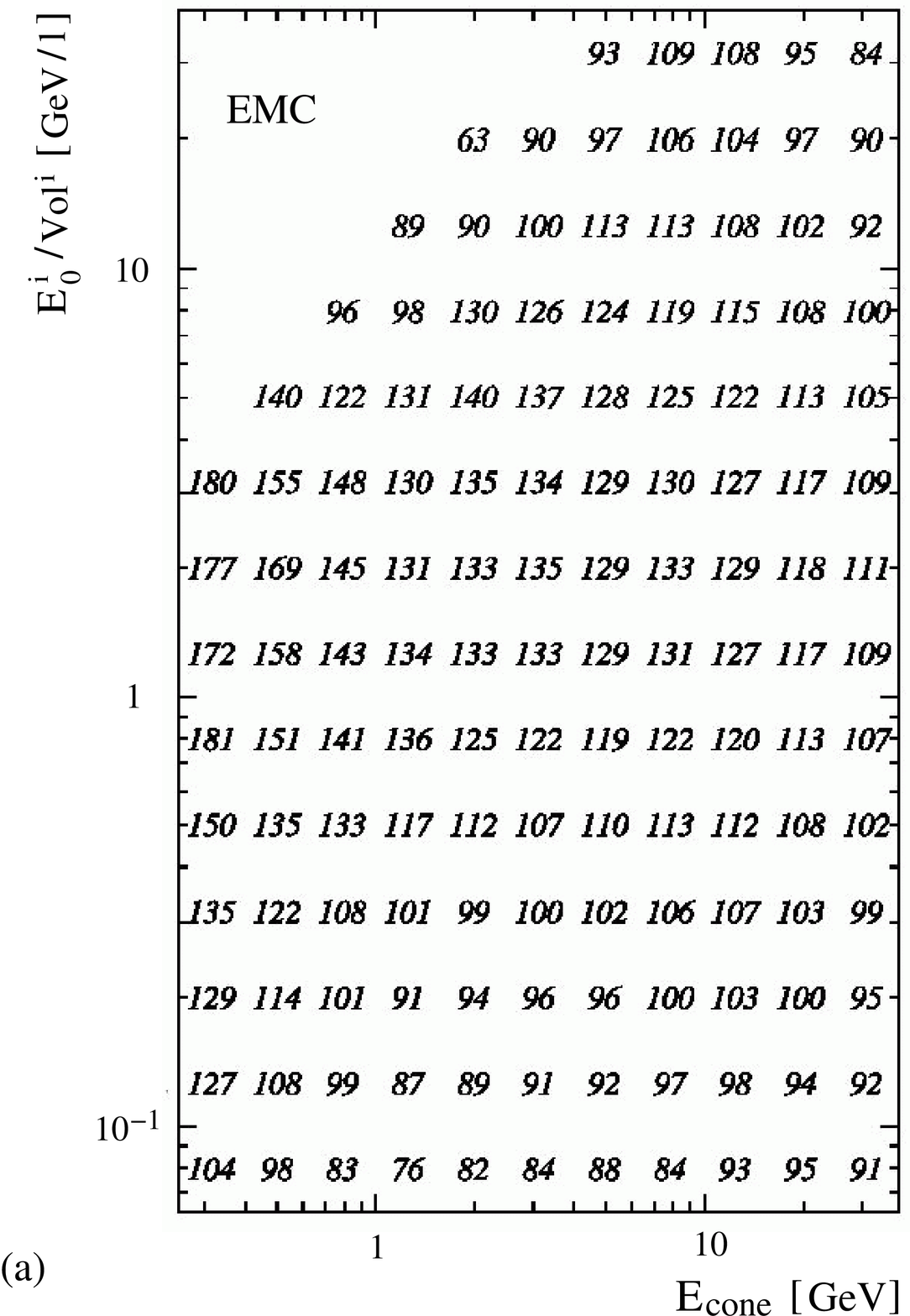}
 \includegraphics[angle=0, width=7cm]{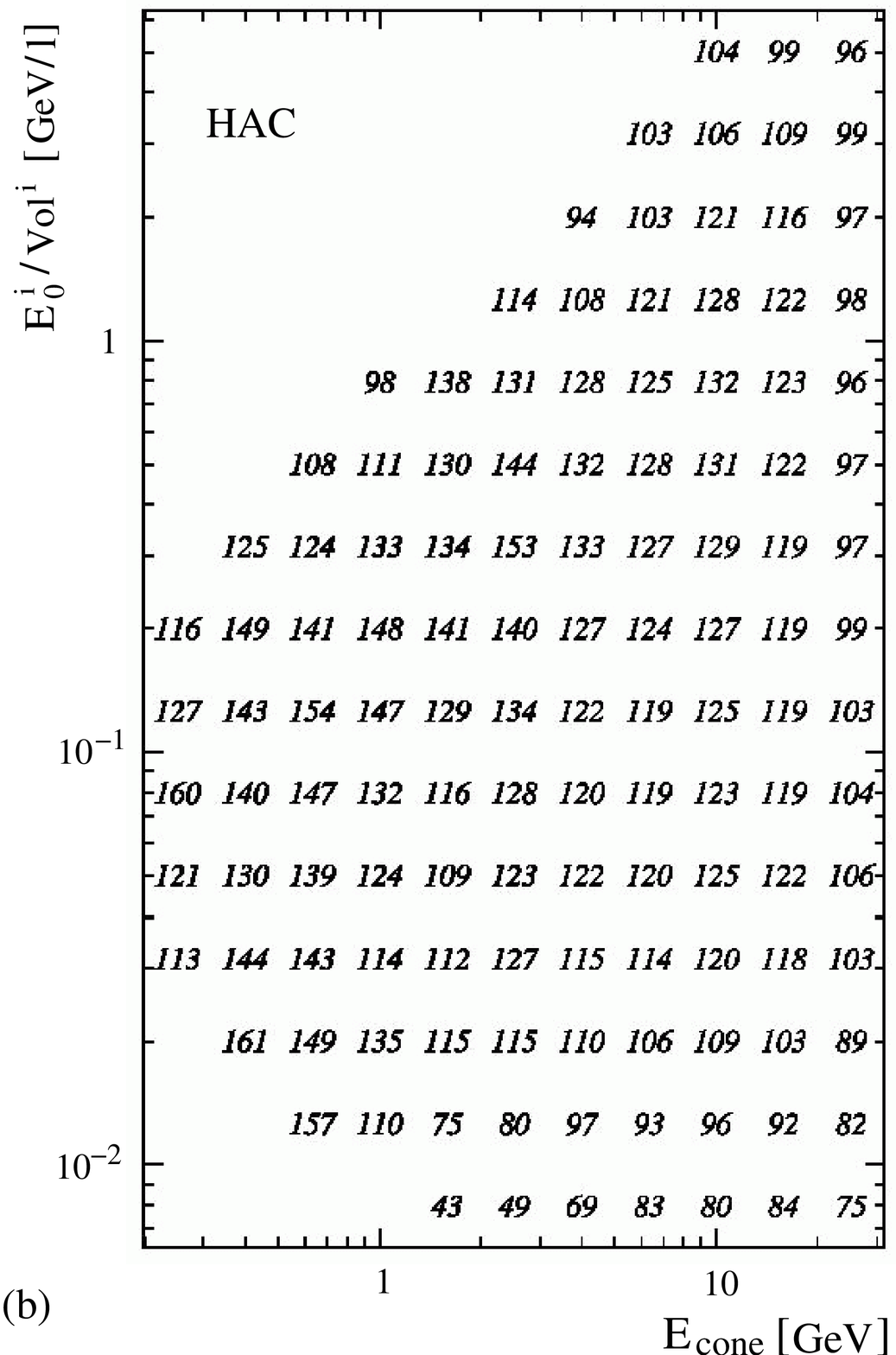}
\end{center}
  \caption{Weighting factors in percent a) for the electromagnetic  and b) the hadronic FB2-modules
of the H1 LAr test calorimeter set-up \cite{Korn} as function of
he energy in the "most energetic cone" and the energy density in
cell $i$.}
\end{figure}

\begin{figure}[h!]
  \begin{center}
   \includegraphics[angle=0, width=8cm]{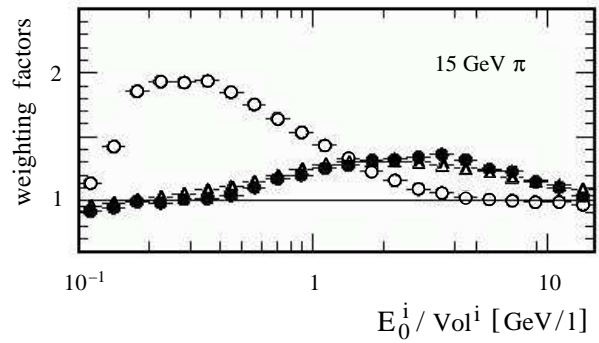}
    \end{center}
 \caption{Comparison of the theoretical weights (full circle) according to
 eq.\ref{eq12}
with the new (open triangle) and the H1 standard weights (open
circle) as function of the energy density in a cell for $15~GeV$
pions.}
 \label{abb11}
\end{figure}

The application of the new weighting algorithm allows to
reconstruct the deposited energy on the cell level. This is
demonstrated by fig.\ref{abb11}, where the theoretical weights
$w^{i}_{th}$ of (eq.\ref{eq12}) are compared to the applied
weights of the H1 standard and of the new algorithm as described
in this paper. While the theoretical (full circles) and the
applied weights of the new algorithm (open triangle) coincide in
good approximation for all values of $\rho_{i}$, the H1 standard
weights (open circles) differ in their trend from the expectation.
Note however that by construction the total energy deposited in
the calorimeter is successfully reconstructed by both algorithms.
The standard H1 weighting procedure achieves this through the
    iterative application of weighting functions according to equation:
$E^i_{rec} = {C_1 exp(-C_2 \frac{E^i_0}{Vol^i}) + C_3}_{EMC/HAC}
\cdot E^i_0$, where $C_i$ are parameter functions, which are
depending on the jet
     energy and polar angle of the jet \cite{H12}.
     These weighting and parameter functions have been determined in an
     iterative process optimizing both the reconstruction of the total
     energy
     deposited in the hadronic shower and the resolution. In this
     iterative process no emphazis is put on the reconstruction of the
     deposited energy on the level of single cells.

\subsection{Correction of Energy Losses due to Noise Cuts}

Only those cells can be considered by the weighting algorithm
which have a finite signal after the noise cuts, i.e the new
weights derived only optimize the energy reconstruction for cells
with a signal above the noise cuts. Hence the losses due to this
cut have to be taken into account in a separate step. In contrast
to this procedure in the H1 standard analysis the noise correction
is included in the weighting factor.

In the new algorithm a first order approximation noise correction
is derived using information of cells belonging to the "most
energetic cone". All cells of this cone are considered including
those which do not belong to a hadronic cluster. From
 simulated data including detector noise the total energy deposited in the cone, $E^{dep,
tot}_{cone}$, and the deposited energy of cells passing the noise
cut $E^{dep, rem}_{cone}$ are calculated. The correction is given
by the expression

\begin{equation}
\Delta E^{dep, cor}_{cone} = E^{dep, tot}_{cone} - E^{dep,
rem}_{cone}. \label{eq17}\end{equation}

\begin{figure}[h!]
  \begin{center}
  \includegraphics[angle=0, width=8cm]{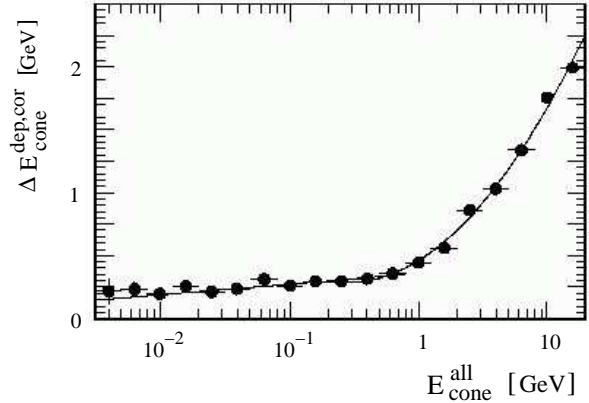}
    \end{center}
 \caption{Correction defined in eq.(\ref{eq17}) considering the losses due to the
noise cuts as a function of the sum of all cells in the "most
energetic cone".}
 \label{abb6}
\end{figure}

The energy correction is evaluated as a function of

\begin{equation}
E^{all}_{cone} = \sum_{j}~E_{rec}^{j} \label{eq18}\end{equation}

on the electromagnetic scale, where the sum runs over all cells in
the "most energetic cone". It is added to the weighted energy of
the cone. The result is shown in fig.\ref{abb6}. The weak energy
dependence for $E^{all}_{cone} \leq 1~GeV$ is due to the fact that
in cones with small energy the fractional noise contribution is
large hence the signal loss due to the noise cuts is small. With
increasing $E^{all}_{cone}$, i.e. increasing shower energy, more
cells in the cone have a signal which can be suppressed by the
noise cut, hence the influence of the noise cuts grows.

\section{Comparison with Simulation and Test Beam Data}

The new algorithm is applied to a set of simulated data. The
geometry of the simulation corresponds to the CERN test beam
configuration \cite{Korn} with an impact angle $\theta =
33.73^{o}$ of the primary pion. Also the geometry of the detector
and the cryostat are adapted to the CERN test set-up \cite{Korn}.

\begin{figure}[h!]
  \begin{center}
   \includegraphics[angle=0, width=13cm]{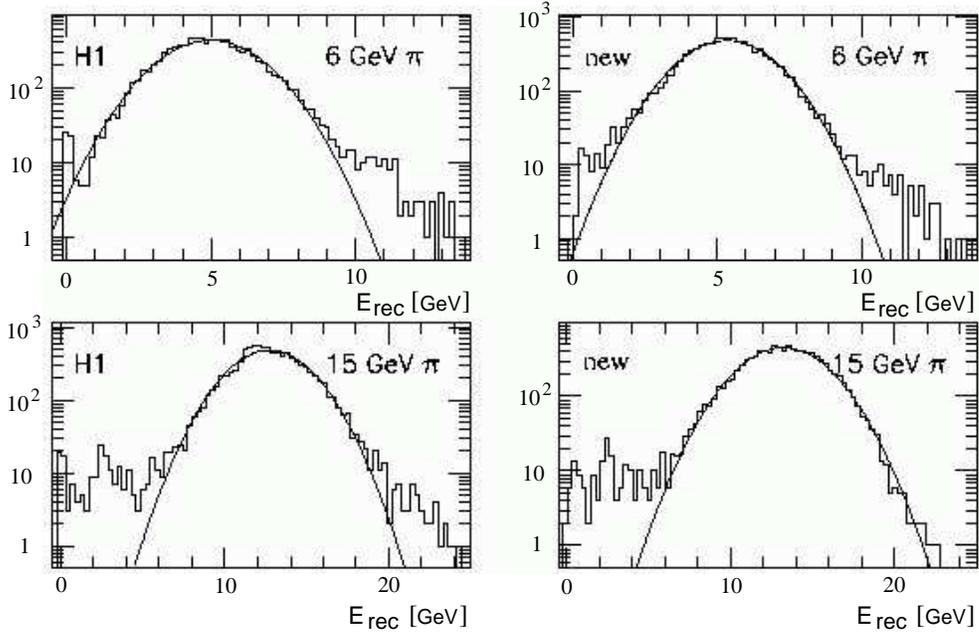}

    \end{center}
 \caption{Comparison of the reconstructed energy distributions a) applying the standard H1
(left row) and b) the new weights (right row) for $\pi$-mesons of
two energies of the primary pion (Monte Carlo simulation).}
 \label{abb7}
\end{figure}

Typical distributions of the reconstructed energy are shown in
fig.\ref{abb7}a for the standard H1 and the new improved algorithm
(fig.\ref{abb7}b). For both weighting procedures a Gaussian shape
of the energy distribution is achieved in the peak region. For the
new algorithm the tails are smaller, especially at higher beam
energies the high energy tail is strongly suppressed. Note that
energy losses due to material in front of the calorimeter are not
corrected for, hence the reconstructed and the primary energy
differ.

\begin{figure}[h!]
  \begin{center}
   \includegraphics[angle=0, width=12cm]{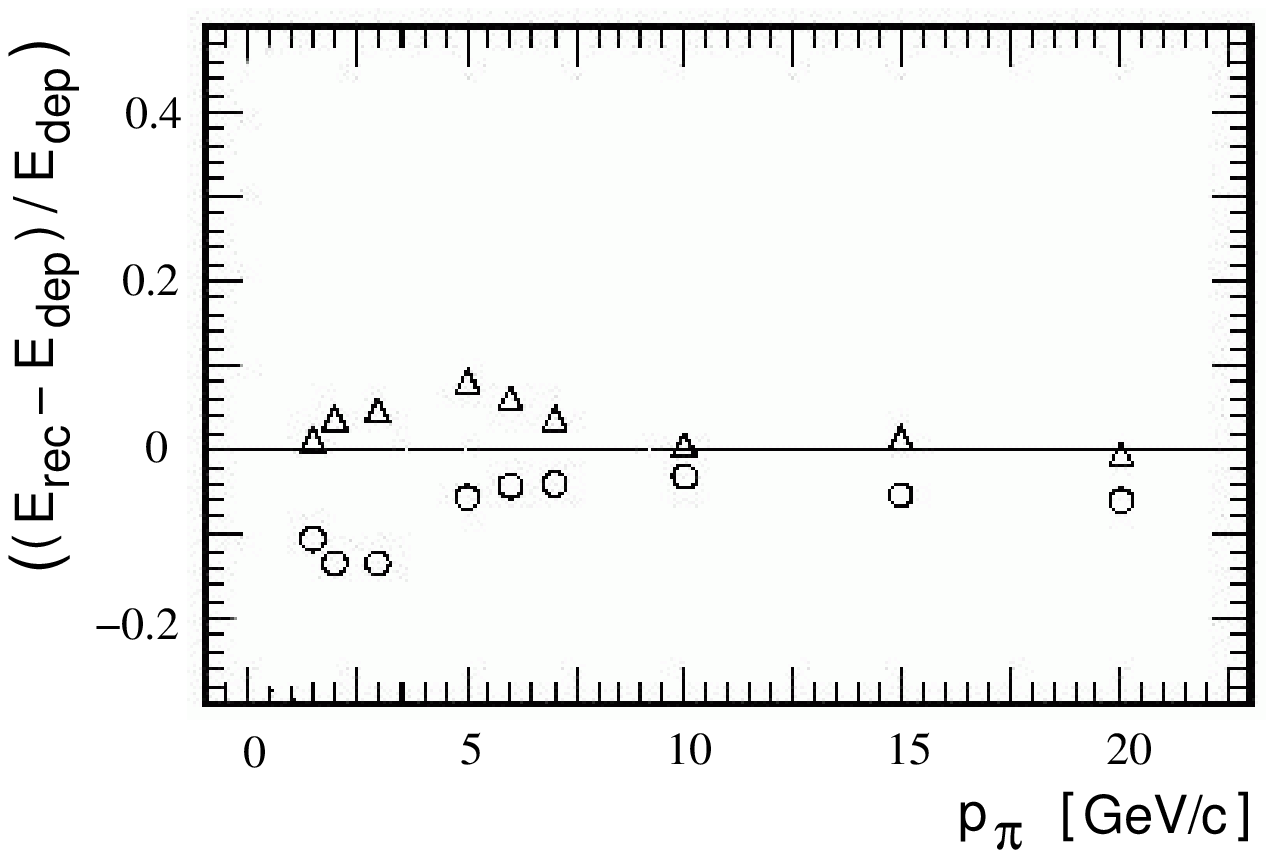}
    \end{center}
 \caption{Difference between reconstructed and deposited energy normalized to the
deposited energy as function of the momentum of the primary pion
for the H1 standard (open circle) and the new weights (triangle)
(Monte Carlo simulation).}
 \label{abb8}
\end{figure}

\begin{figure}[h!]
  \begin{center}
   \includegraphics[angle=0, width=12cm]{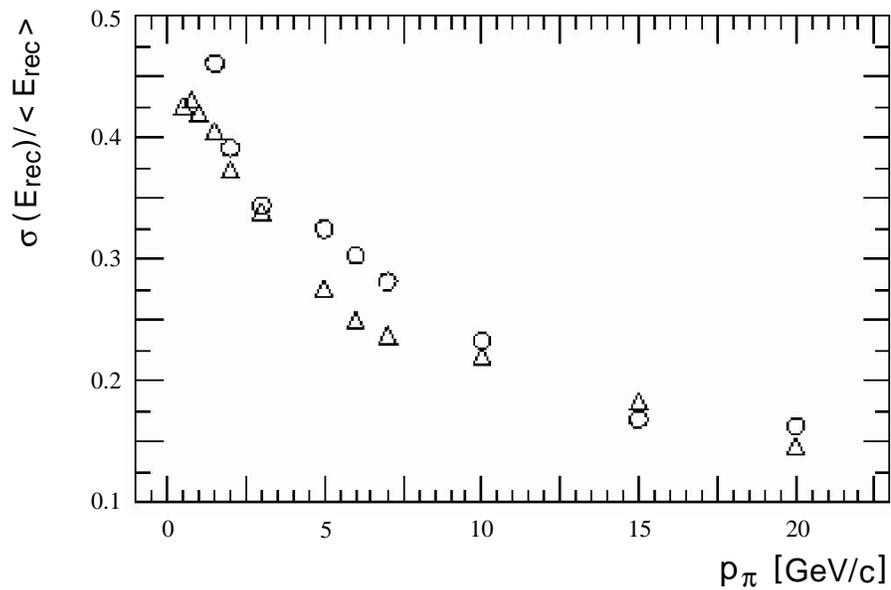}
    \end{center}
 \caption{Relative energy resolution of the reconstructed energies for H1
 standard (circles)
and for new weights (triangles) (Monte Carlo simulation).}
 \label{abb9}
\end{figure}

The energy response is linear within $2 \%$ for energies above $2
~ GeV$ if one uses the new algorithm while for the standard H1
procedure larger deviation from linearity are observed
(fig.\ref{abb8}). An even better linearity, especially in the
region of 5GeV, would be
 achievable when using a finer grid in the Monte Carlo simulations
 for the determination of the new weighting factors.
 The use of the new algorithm furthermore leads to an improved
 energy resolution of the calorimeter (fig.\ref{abb9}).

\begin{figure}[h!]
  \begin{center}
   \includegraphics[angle=0, width=8.5cm]{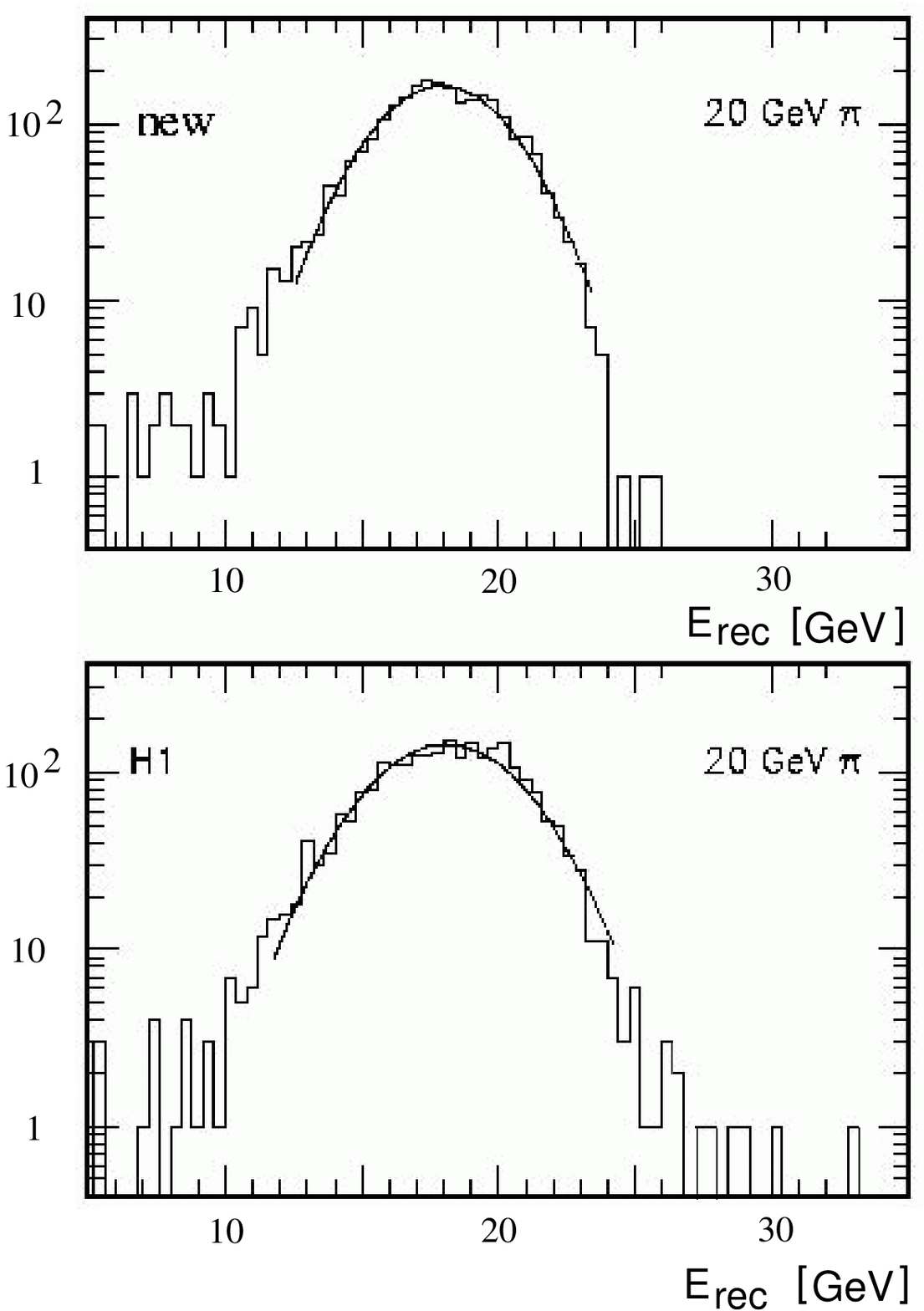}
    \end{center}
 \caption{Distribution of reconstructed energies for test beam data collected at the CERN
 test beam H6,
  data were collected with test
modules of the H1 LAr calorimeter: reconstructed a) with new and
b) with H1 standard weights.}
 \label{abb10}
\end{figure}

Finally the new algorithm also was applied to real data taken at
the CERN SPS test beam H6 with a pion beam energy of $20~GeV$. In
fig.\ref{abb10} the reconstructed energy distribution for the two
algorithms are compared, again showing an improvement when using
the new weigths confirming the results from the study of simulated
data.

\section{Conclusion}

A new weighting algorithm has been developed to correct the
different response of the H1 LAr calorimeter to electromagnetic
and hadronic showers. The essential difference between the new and
the standard H1 algorithm consists in the separate treatment of
weight factors and the corrections for noise cuts. This separation
allows to reconstruct the energy deposited in the cells of the
calorimeter properly. Hence the new algorithm described in this
paper not only improves the linearity and the energy resolution
(figs.\ref{abb8}, \ref{abb9}) but in contrast to the H1 standard
algorithm it allows to reproduce the shower shape
(fig.\ref{abb11}).

Recently this algorithm has been generalized by J.Marks
\cite{Marks} in such a way, that it can be applied to all modules
of the H1 LAr calorimeter at the HERA ep storage ring. Moreover
improved corrections for the noise cuts were developed.

\newpage

{\bf\underline {Acknowledgements}}

This work is based on the untiring work of many members of the H1
calorimeter group in the years 1986 - 1996 to develop a calibration
procedure and a version of the weighting algorithm which
successfully was applied in the last 10 years by the H1
Collaboration in their data analysis. Special thanks are due to Drs.
J. Gayler, P. Loch and J. Spiekermann for stimulating discussions
and for help and advice and Dr. H.-C. Schultz-Coulon for careful and
critical reading of the manuscript. This work was supported by the
BMBF, Bonn under contract number 05 H11PEA/6. \c{C}i\v{g}dem
\.{I}\c{s}sever thanks the Studienstiftung des Deutschen Volkes for
a fellowship and the University of Dortmund for a special research
position. Kerstin Borras was supported by a Lise Meitner-fellowship
of the MWF, Düsseldorf.

\end{document}